\documentclass[preprint,12pt]{elsarticle}

\usepackage{amsmath}
\usepackage{graphicx}
\usepackage{siunitx}
\usepackage{url}
\usepackage{lineno}
\usepackage{physics2}

\usephysicsmodule{ab,ab.braket}

\journal{Optics and Lasers in Engineering}

\begin{document}

\begin{frontmatter}
  \title{Parallel Phase-shifting Digital Ghost Holography}
  \author{Shuhei Yoshida}
  \ead{yshuhei@ele.kindai.ac.jp}
  \affiliation{organization={Department of Electrical, Electronic and Communication Engineering, Kindai University},
    addressline={3-4-1 Kowakae},
    city={Higashiosaka},
    postcode={577-8502},
    state={Osaka},
    country={Japan}}
    
    \begin{abstract}
      The ghost imaging (GI) technique, which has attracted attention as a highly sensitive and noise-resistant technique, employs a spatially modulated illuminating light and a single-pixel detector. Generally, the information acquired by GI is the transmittance or reflectance distribution of an object. A method has also been proposed to measure the complex amplitude by applying digital holography (DH) techniques. These methods irradiate phase-modulated illuminating lights onto an object, and the intensities of the interference lights between the lights interacting with the object and the reference light are measured. Then, the complex amplitude of the object light is reconstructed based on the correlation between the light intensities and the phase patterns. In DH-based GI, it is necessary to remove unwanted components from the interferogram by phase shifting, which requires more measurements than the conventional GI method. Thus, we propose a technique to reconstruct the complex amplitude in DH-based GI without increasing the number of measurements using parallel phase-shifting optics. In the proposed method, interferograms phase-shifted in steps of $\pi/2$ with waveplates are divided into four using polarization beam splitters (PBS), and their intensities are measured simultaneously. The object light component can be extracted from the intensities of these four interferograms. We demonstrate the effectiveness of the proposed method through experiments.
    \end{abstract}



\end{frontmatter}


\section{Introduction}
In the mid-1990s, ghost imaging (GI), which employs a single-pixel detector, e.g., a photodiode (PD), was proposed for acquiring images of objects \cite{Pittman1995,Strekalov1995,Bennink2002,Gatti2004,Ferri2005,Shapiro2008,Bromberg2009}, and it has since attracted attention because of its high sensitivity and noise immunity \cite{Meyers2011,Shibuya2015}, as well as its ability to detect wavelength bands where image sensor integration and sensitivity are not sufficient \cite{Chan2008,Xu2010,Watts2014,Radwell2014,Gibson2017}. In GI using classical light \cite{Bennink2002,Gatti2004,Ferri2005,Shapiro2008,Bromberg2009}, spatially modulated illuminating lights are irradiated onto an object, and the single-pixel detector measures the intensities of the lights interacting with the object. In this process, the light intensities are measured multiple times while changing the illumination pattern, and the object image is reconstructed based on the correlation between the measured light intensities and the illumination patterns. The single-pixel imaging (SPI) imaging technique is similar to GI. In SPI, the measurement process is handled as a linear transformation, and the imaging is performed by applying compressed sensing \cite{Duarte2008,Katz2009,Gibson2020}. Both GI and SPI have been applied effectively in various fields, e.g., bioimaging \cite{Studera2012,Zhao2023apr}, 3D imaging \cite{Zhang2016,Sun2016,Sun2019}, terahertz imaging \cite{Chan2008,Xu2010,Watts2014,Olivieri2020,Valls2020,Zanotto2020,Ismagilov2022,Deng2023,Liu2023,Guan2023}, multispectral and hyperspectral imaging \cite{Studera2012,Bian2016,Li2017,Jin2017,Pian2017,Olivieri2020,Jiang2022,Zhao2023nov,Zhang2024,Zuljevic2024,Wenwen2024} scattering imaging \cite{Tajahuerce2014,Durn2015,Jauregui-Snchez2019,Soltanlou2019,Lin2022mar,Gao2022,Xiao2022,Peng2023,Lin2023,Hao2023,Hao2024,Wang2024,Hao2025}, underwater imaging \cite{Yang2021,Feng2023,Li2023,Gao2024,Hu2024,Feng2025}, image classification \cite{Zhang2020,Cao2021,Hanawa2022,Yao2024,He2024,Yang2025}, remote sensing \cite{Lai2023,Yin2025}, optical cryptography \cite{Zhao2020,Zheng2021,Lin2022sep,Zheng2022,Liu2022,Kang2023}, and polarization imaging \cite{Durn2012,Soldevila2013,Shi2014,Kim2020,Seow2020}. Generally, the information obtained by GI and SPI is an object's transmittance or reflectance distribution; however, methods to measure the complex amplitude, including the phase, have been proposed by applying digital holography (DH) \cite{Clemente2012,Clemente2013,Martnez-Len2017,Gonzlez2018}. In digital ghost holography (DGH) or single-pixel digital holography (SPDH), phase-modulated illuminating lights via a spatial light modulator (SLM) are irradiated onto the target object, and the intensities of the interference lights between the lights that interact with the object and the reference light are measured. Then, the complex amplitude of the object light is obtained by correlating the phase patterns irradiated on the object with the intensities of the interfering lights.

However, compared to GI and SPI, DGH and SPDH have two problems that reduce their availability, i.e., the complexity of the optical system and the increased number of measurements required to remove unwanted light. Two-beam interferometers, e.g., Mach-Zehnder \cite{Clemente2012,Clemente2013,Gonzlez2018,Endo2022} or Michelson interferometers \cite{Martnez-Len2017,Endo2019}, are frequently employed as optical systems for DGH and SPDH; however, the optical systems become more complex and less stable. Also, in addition to the object light as in DH, the interferograms obtained by DGH and SPDH contain unwanted components, e.g., zeroth-order and conjugate light; thus, a phase-shifting method \cite{Yamaguchi1997} must be employed to remove such components. The phase-shifting method adds bias phases to the illumination phase pattern as $0, \pi/2, \pi, 3\pi/2$, etc., and measures the interference light intensities. In addition, certain operations extract the object light component from the light intensities; thus, DGH and SPDH require several (at least three) more measurements compared with the conventional GI and SPDH methods. Various methods that utilize common-path optics have been proposed to address the stability degradation caused by the complexity of the optical system \cite{Shin2018apr,Shin2018oct,Liu2018,Ota2018,Zhao2019}. The use of common-path optics in these methods reduces the optical complexity and instability problems associated with two-beam interferometers. However, to the best of the author's knowledge, only a few efforts have attempted to increase the number of measurements. Yoneda et al. proposed the common-path off-axis single-pixel holographic imaging (COSHI) method \cite{Yoneda2022}, which does not require phase-shifting. The COSHI method reconstructs off-axis holograms containing the object light using Hadamard bases superimposed on the linear phase as the illuminating light. Then, Fourier fringe analysis \cite{Takeda1982} is applied to the holograms to obtain the complex amplitude of the object light. The COSHI method can obtain the complex amplitude, including the phase, without increasing the number of measurements compared with the GI and SPI methods. Thus, the COSHI method is useful; however, the frequency filter limits the spatial bandwidth product and reduces the resolution.

This paper proposes a method to reconstruct the complex amplitude of the object light by applying the parallel four-step phase-shifting method \cite{Kakue2010} to the DGH method without increasing the number of measurements, and we demonstrate the effectiveness of the proposed method through experiments. In the parallel four-step phase-shifting method used in DH, the spatially superimposed phase-shifted holograms are separated by a polarization beam splitter (PBS) or polarization imaging camera and acquired simultaneously. In the proposed method, interferograms phase-shifted in steps of $\pi/2$ with waveplates are divided into four using PBSs, and their intensities are measured. Using balanced detectors effectively removes the zeroth-order light component. Note that the proposed method requires a two-beam interferometer; however, it can measure the complex amplitude of the object light with the same number of measurements and resolution as the conventional GI and SPI methods. Note that the resources used in this study are publicly available through GitHub \cite{GitHub}.

\section{Principle}
In the proposed method, a horizontally polarized illuminating pattern is incident on the object, and the transmitted light becomes the object light. Horizontally polarized object light and vertically polarized reference light are input to a beam splitter (BS), and the two outputs of the BS are phase-shifted by a quarter-wave plate (QWP) and half-wave plate (HWP), respectively. Here, the lights transmitted through the QWP and HWP are separated by horizontal and vertical polarization using PBSs to obtain four interferograms whose phases are shifted in steps of $\pi/2$. Note that the object light component can be extracted from the power of these four interferograms; thus, the complex amplitude can be reconstructed with the same number of measurements as the GI method. The principle of the proposed method is explained in the following.

\begin{figure}[!ht]
  \centering
  \includegraphics[scale=0.5]{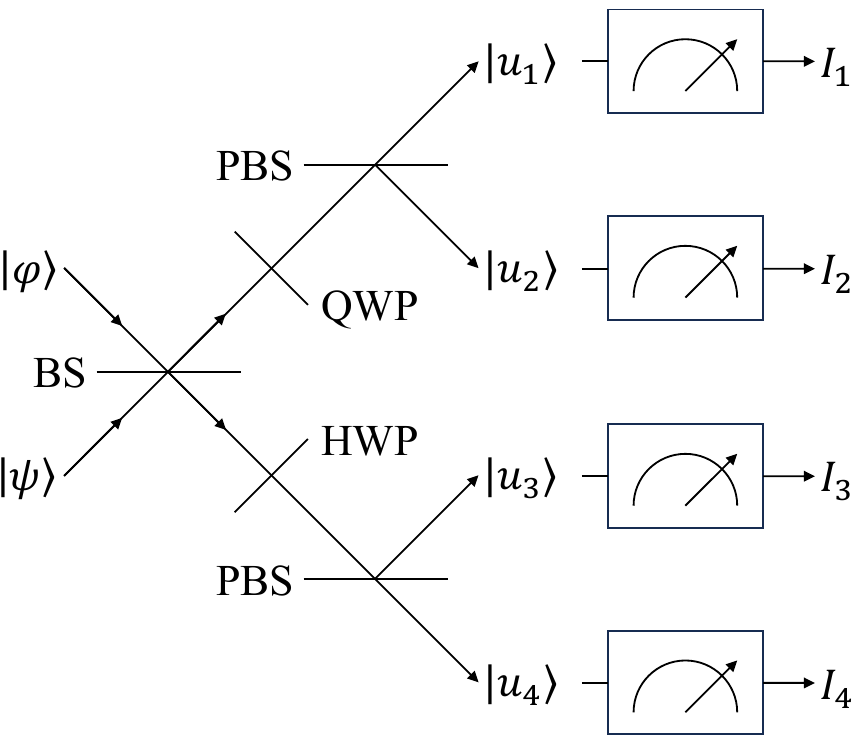}
  \caption{Relationship between $\ket|\varphi>$, $\ket|\psi>$ and $I_{1},\dots,I_{4}$. Here, we assume that the BS is symmetric for both the input and output. The angles of QWP and HWP are the inclination of the fast axis to the horizontal axis.}
  \label{fig:system}
\end{figure}

The complex amplitude distribution of the sample is $\alpha$, and the phase pattern displayed on the SLM is $\phi_{n}~(n=1,2,\dots)$. Then, the illuminating light generated by the SLM passes through the sample to produce the object light $\ket|\varphi>$ as follows:
\begin{equation}
  \ket|\varphi> = \ket|\mathrm{e}^{\mathrm{i}\phi_{n}}\alpha>\ket|H>,
\end{equation}
where $\ket|H>$ denotes the horizontal polarization basis. The reference light $\ket|\psi>$ is a plane wave with vertical polarization and is expressed as follows:
\begin{equation}
  \ket|\psi> = \ket|\beta>\ket|V>,
\end{equation}
where $\beta$ denotes the complex amplitude, and $\ket|V>$ denotes the vertical polarization basis. Then, the object light $\ket|\varphi>$ and reference light $\ket|\psi>$ enter the BS. The transfer matrix of the BS symmetric for the input and output is given as follows \cite{Loudon2000}:
\begin{equation}
  \mathbf{T}_{\mathrm{BS}} = \frac{1}{\sqrt{2}}
    \begin{pmatrix}
      1 & \mathrm{i}\\
      \mathrm{i} & 1\\
    \end{pmatrix}.
\end{equation}
In addition, the output of the BS expressed as follows:
\begin{equation}
  \mathbf{T}_{\mathrm{BS}}
    \begin{pmatrix}
      \ket|\varphi>\\
      \ket|\psi>\\
    \end{pmatrix} = \frac{1}{\sqrt{2}}
      \begin{pmatrix}
        \ket|\varphi> + \mathrm{i}\ket|\psi>\\
        \mathrm{i}\ket|\varphi> + \ket|\psi>\\
      \end{pmatrix} =
        \begin{pmatrix}
          \ket|\mathrm{e}^{\mathrm{i}\phi_{n}}\alpha>\ket|H> + \mathrm{i}\ket|\beta>\ket|V>\\
          \mathrm{i}\ket|\mathrm{e}^{\mathrm{i}\phi_{n}}\alpha>\ket|H> + \ket|\beta>\ket|V>\\
        \end{pmatrix}.
\end{equation}
Passing each output of the BS through the QWP and HWP yields the following, respectively:
\begin{align}
  \mathbf{J}_{\mathrm{QWP}}\ab({\ang{45}})\ab\{\frac{1}{\sqrt{2}}\ab(\ket|\varphi> + \mathrm{i}\ket|\psi>)\} &= \frac{1}{2}\ab\{\ket|\mathrm{e}^{\mathrm{i}\phi_{n}}\alpha>\ab(\ket|H> - \mathrm{i}\ket|V> + \ket|\beta>\ab(\ket|H> + \mathrm{i}\ket|V>))\},
  \label{eq:qwp}\\
  \mathbf{J}_{\mathrm{HWP}}\ab({\ang{22.5}})\ab\{\frac{1}{\sqrt{2}}\ab(\mathrm{i}\ket|\varphi> + \ket|\psi>)\} &= \frac{1}{2}\ab\{\ket|\mathrm{e}^{\mathrm{i}\phi_{n}}\alpha>\ab(\ket|H> + \ket|V> + \mathrm{i}\ket|\beta>\ab(\ket|V> - \ket|H>))\}.
  \label{eq:hwp}
\end{align}
Here, $\mathbf{J}_{\mathrm{QWP}}(\theta)$ and $\mathbf{J}_{\mathrm{HWP}}(\theta)$ denote the QWP and HWP with the fast axis tilted $\theta$ from the horizontal axis, respectively. $\mathbf{J}_{\mathrm{QWP}}(\theta)$ and $\mathbf{J}_{\mathrm{HWP}}(\theta)$ are expressed by the Jones matrix as follows \cite{Saleh2007}.
\begin{subequations}
  \begin{align}
    \mathbf{J}_{\mathrm{QWP}}(\theta) &= \frac{1}{\sqrt{2}}
      \begin{pmatrix}
        1 - \mathrm{i}\cos{2\theta} & -\mathrm{i}\sin{2\theta}\\
        -\mathrm{i}\sin{2\theta} & 1 + \mathrm{i}\cos{2\theta}\\
      \end{pmatrix},\\
    \mathbf{J}_{\mathrm{HWP}}(\theta) &= -\mathrm{i}
      \begin{pmatrix}
        \cos{2\theta} & \sin{2\theta}\\
        \sin{2\theta} & -\cos{2\theta}\\
      \end{pmatrix}
  \end{align}
\end{subequations}
When transmitted through the QWP and HWP, the light in Eqs. (\ref{eq:qwp}) and (\ref{eq:hwp}) is divided into horizontal and vertical components, respectively, by the PBS and $u_{1},\dots,u_{4}$, which are expressed as follows:
\begin{subequations}
  \begin{align}
    \ket|u_{1}> &= \frac{1}{2}\ket|H>\ab(\ket|\mathrm{e}^{\mathrm{i}\phi_{n}}\alpha> + \ket|\beta>),\\
    \ket|u_{2}> &= -\frac{\mathrm{i}}{2}\ket|V>\ab(\ket|\mathrm{e}^{\mathrm{i}\phi_{n}}\alpha> - \ket|\beta>),\\
    \ket|u_{3}> &= \frac{1}{2}\ket|H>\ab(\ket|\mathrm{e}^{\mathrm{i}\phi_{n}}\alpha> - \mathrm{i}\ket|\beta>),\\
    \ket|u_{4}> &= \frac{1}{2}\ket|V>\ab(\ket|\mathrm{e}^{\mathrm{i}\phi_{n}}\alpha> + \mathrm{i}\ket|\beta>).
  \end{align}
\end{subequations}
Here, $I_{1},\dots,I_{4}$ denote the light intensities of $u_{1},\dots,u_{4}$, which are expressed as follows:
\begin{subequations}
  \begin{align}
    I_{1} &= \braket<u_{1}|u_{1}> = \frac{1}{4}\ab(\braket<\alpha|\alpha> + \braket<\beta|\beta> + \braket<\mathrm{e}^{\mathrm{i}\phi_{n}}\alpha|\beta> + \braket<\beta|\mathrm{e}^{\mathrm{i}\phi_{n}}\alpha>),\\
    I_{2} &= \braket<u_{2}|u_{2}> = \frac{1}{4}\ab(\braket<\alpha|\alpha> + \braket<\beta|\beta> - \braket<\mathrm{e}^{\mathrm{i}\phi_{n}}\alpha|\beta> - \braket<\beta|\mathrm{e}^{\mathrm{i}\phi_{n}}\alpha>),\\
    I_{3} &= \braket<u_{3}|u_{3}> = \frac{1}{4}\ab(\braket<\alpha|\alpha> + \braket<\beta|\beta> + \mathrm{i}\braket<\mathrm{e}^{\mathrm{i}\phi_{n}}\alpha|\beta> - \mathrm{i}\braket<\beta|\mathrm{e}^{\mathrm{i}\phi_{n}}\alpha>),\\
    I_{4} &= \braket<u_{4}|u_{4}> = \frac{1}{4}\ab(\braket<\alpha|\alpha> + \braket<\beta|\beta> - \mathrm{i}\braket<\mathrm{e}^{\mathrm{i}\phi_{n}}\alpha|\beta> + \mathrm{i}\braket<\beta|\mathrm{e}^{\mathrm{i}\phi_{n}}\alpha>).
  \end{align}
\end{subequations}
The relationship between the input $\ket|\varphi>$, $\ket|\psi>$ to the BS and intensities $I_{1},\dots,I_{4}$ is shown in Fig. \ref{fig:system}. Note that the interferogram phases are shifted in steps of $\pi/2$; thus, the complex amplitude $\alpha$ can be reconstructed using the following procedure.
Using $I_{1},\dots,I_{4}$, we obtain the following relationship.
\begin{equation}
  \ab(I_{1} - I_{2}) + \mathrm{i}\ab(I_{3} - I_{4}) = \braket<\beta|\mathrm{e}^{\mathrm{i}\phi_{n}}\alpha> = \braket<\mathrm{e}^{-\mathrm{i}\phi_{n}}\beta|\alpha> = \braket<\mathrm{e}^{-\mathrm{i}\phi_{n}}|\alpha>
  \label{eq:component}
\end{equation}
In this context, the amplitude of the reference light is assumed to be spatially uniform and $\beta=1$. In Eq. (\ref{eq:component}), $I_{1}-I_{2}$ and $I_{3}-I_{4}$ can be measured using balanced detectors. Here, $\braket<\mathrm{e}^{-\mathrm{i}\phi_{n}}|\alpha>$ is the component of $\ket|\alpha>$ for the basis $\ket|\mathrm{e}^{-\mathrm{i}\phi_{n}}>$; thus, the complex amplitude $\ket|\alpha>$ can be obtained as follows:
\begin{equation}
  \ket|\alpha> = \sum_{n}\braket<\mathrm{e}^{-\mathrm{i}\phi_{n}}|\alpha>\ket|\mathrm{e}^{-\mathrm{i}\phi_{n}}>.
  \label{eq:amplitude}
\end{equation}

\subsection{Spatial orthogonal pattern}
\begin{figure}[!ht]
  \centering
  \includegraphics[scale=0.5]{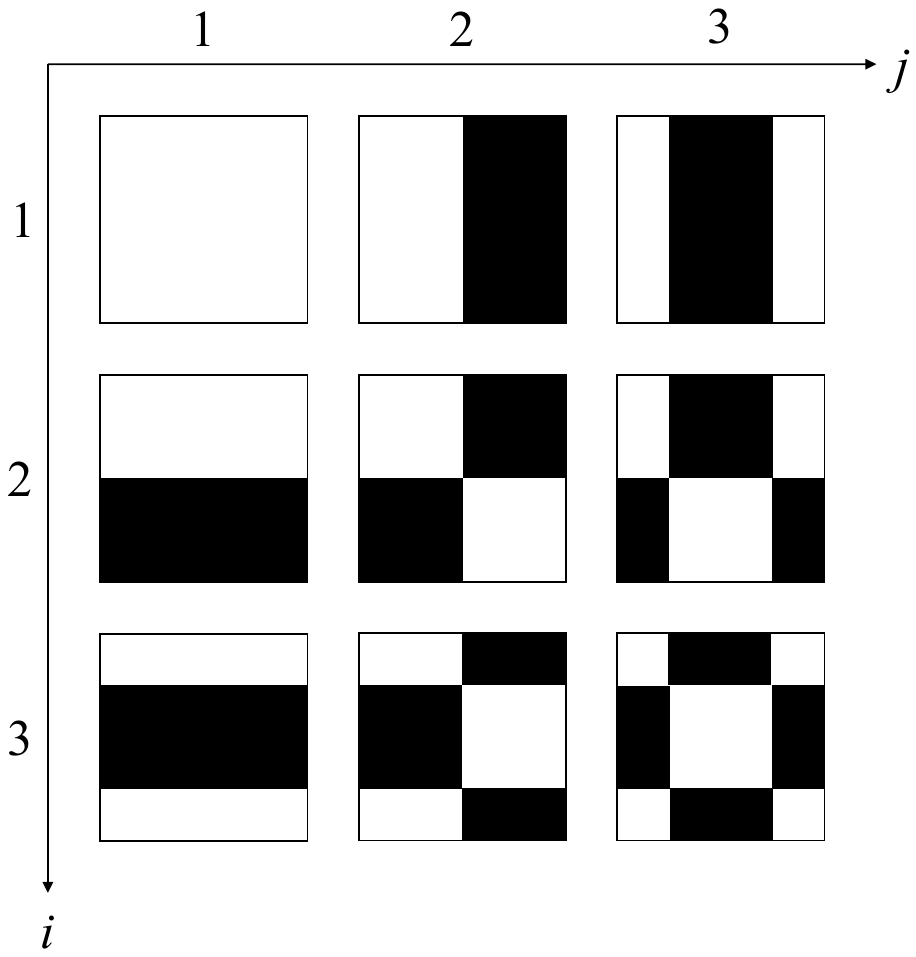}
  \caption{Example of spatial orthogonal pattern $\mathbf{W}_{ij}$, where $+1$ and $-1$ are shown in white and black, respectively.}
  \label{fig:hadamard}
\end{figure}

In this study, spatial orthogonal bases are used as the phase pattern $\phi_{n}$ to measure the spatial frequency without overlap. A common way to construct the spatial orthogonal bases is to use a Hadamard matrix (as shown below), which is an orthogonal matrix with $+1$ or $-1$ elements, and the $2^{k}$-order Hadamard matrix can be obtained recursively as follows \cite{Pratt1969,Shibuya2015}.
\begin{equation}
  \begin{split}
    \mathbf{H}_{2} &=
      \begin{pmatrix}
        1 & 1\\
        1 & -1\\
      \end{pmatrix}\\
    \mathbf{H}_{2^{k}} &=
      \begin{pmatrix}
        \mathbf{H}_{2^{k-1}} & \mathbf{H}_{2^{k-1}}\\
        \mathbf{H}_{2^{k-1}} & -\mathbf{H}_{2^{k-1}}\\
      \end{pmatrix}
      = \mathbf{H}_{2}\otimes\mathbf{H}_{2^{k-1}}
  \end{split}
\end{equation}
Here, $\otimes$ is the Kronecker product. After sorting the rows of the Hadamard matrix in ascending order of spatial frequency, the product of the $i$-th column $\mathrm{col}(\mathbf{H})_{i}~(i=1,2,\dots,2^{k})$ and the $j$-th row $\mathrm{row}(\mathbf{H})_{j}~(j=1,2,\dots,2^{k})$ defines the matrix $\mathbf{W}_{ij}=\mathrm{col}(\mathbf{H})_{i}~\mathrm{row}(\mathbf{H})_{j}$. We obtain the phase pattern $\phi_{n}~(n=1,2,\dots,2^{2k})$ by replacing $+1$ and $-1$ of $\mathbf{W}_{ij}$ with $0$ and $\pi$, respectively. Fig. \ref{fig:hadamard} shows an example of $\mathbf{W}_{ij}$. In the experiment conducted in this study, $(i,j)=(1,1),(2,1),(1,2),(3,1),(2,2),(1,3),\dots$ were measured in ascending order of spatial frequency.

\section{Experiment}
\subsection{Setup}
\begin{figure}[!ht]
  \centering
  \includegraphics[scale=0.6]{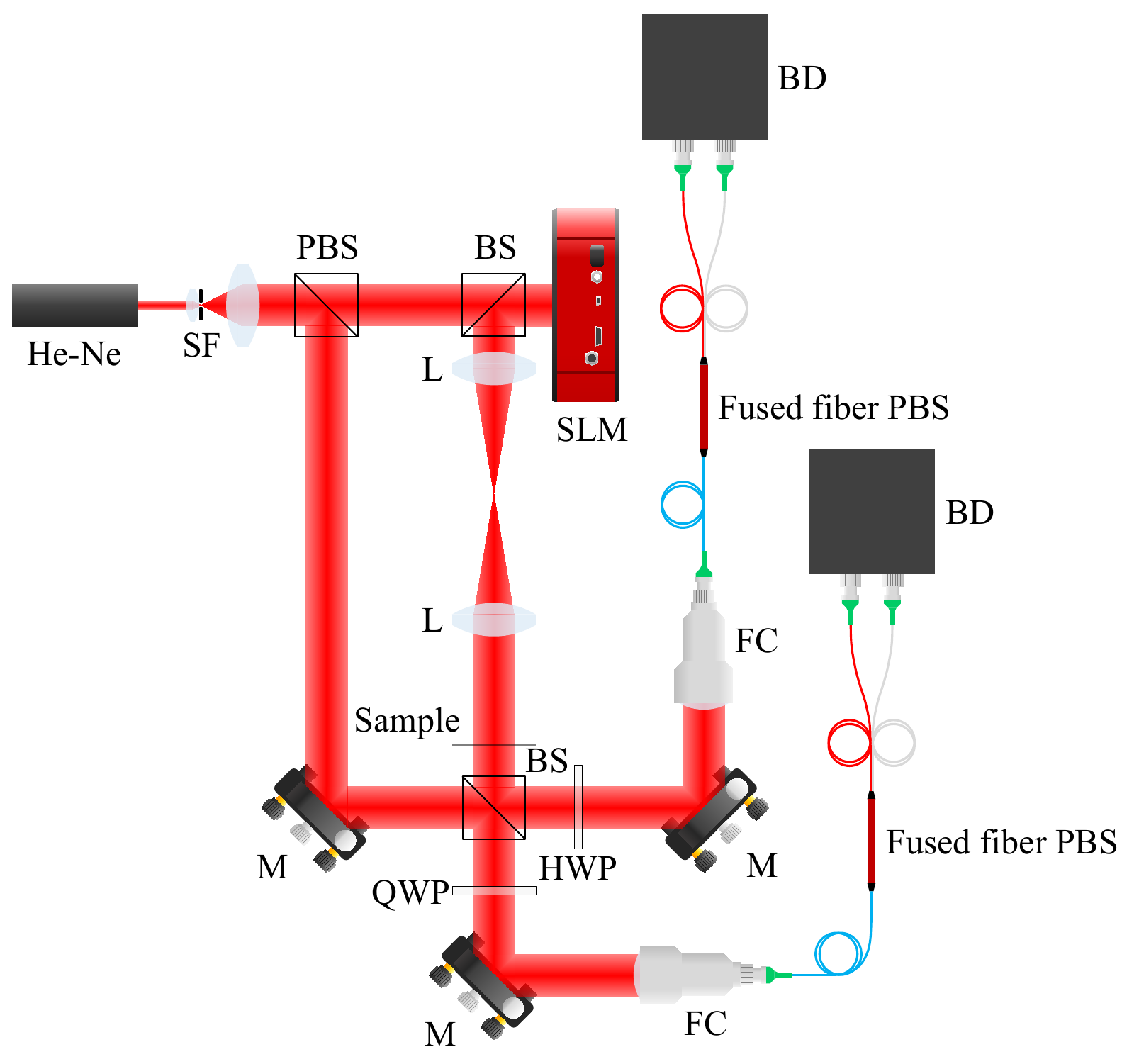}
  \caption{Optics of the proposed method (SF: spatial filter; L: lens; M: mirror; FC: fiber collimator; BD: balanced detector).}
  \label{fig:setup}
\end{figure}

The optical system, i.e., a Mach--Zehnder interferometer, of the proposed method is shown in Fig. \ref{fig:setup}. Here, a He-Ne laser with (wavelength: \SI{632.8}{nm}) was used as the light source, and a liquid crystal SLM (Thorlabs EXULUS-HD2) with a resolution of $1920\times1200$ and a pixel size of \SI{8}{\micro m} was used to generate the phase patterns. The light emitted from the laser was wavefront-shaped by a spatial filter, collimated by a collimating lens, and divided by a PBS. The horizontally polarized component was phase-modulated by the SLM, relayed by the lens system, and transmitted through the sample to become the object light. The vertically polarized component became the reference light without modification. The object light and the reference light merge with the reference light at the BS. In addition, the two outputs of the BS were transmitted through the QWP and HWP, respectively, and then split into orthogonal polarization components using fused fiber PBSs (Thorlabs PFC635A) and input to balanced detectors (Thorlabs PDB450A). The outputs of the two balance detectors are $I_{1}-I_{2}$ and $I_{3}-I_{4}$, respectively. The outputs of the balance detectors were A/D converted, and the wavefront was reconstructed according to Eq. (\ref{eq:amplitude}).

\subsection{Results}
\begin{figure}[!ht]
  \centering
  \includegraphics[scale=0.5]{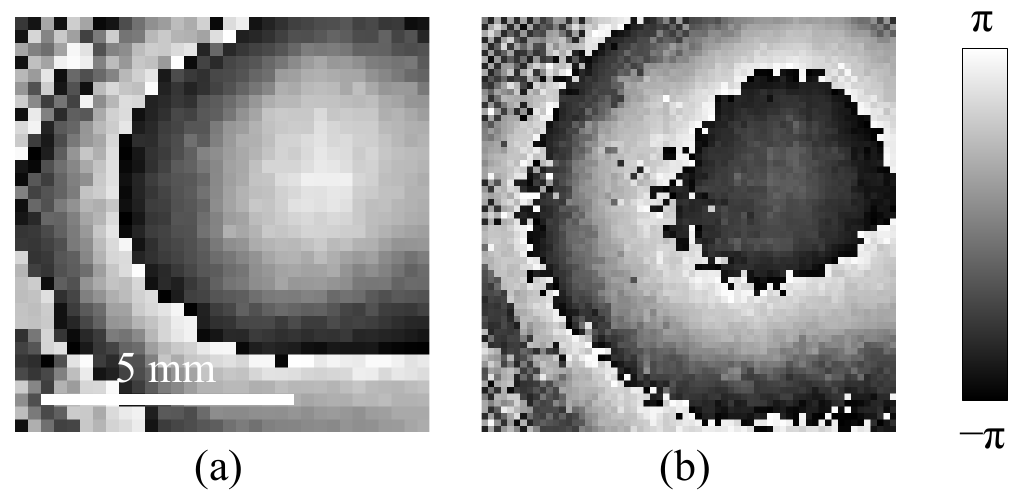}
  \caption{Wavefront measurement results obtained using the proposed method: (a) $32\times32$ resolution and (b) $64\times64$ resolution.}
  \label{fig:wavefront}
\end{figure}

\begin{figure}[!ht]
  \centering
  \includegraphics[scale=0.5]{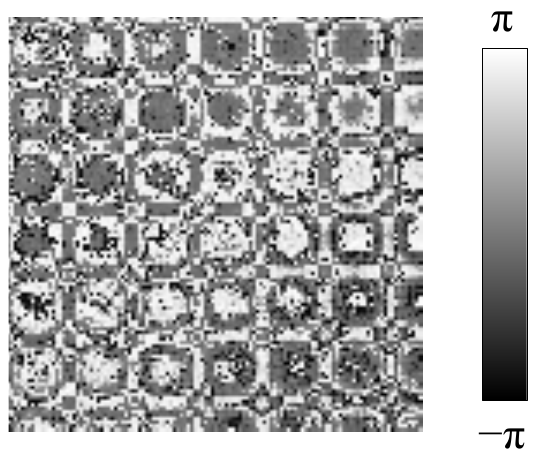}
  \caption{Phase measurement result of the microlens array (resolution: $128\times128$).}
  \label{fig:microlens}
\end{figure}

\begin{figure}[!ht]
  \centering
  \includegraphics[scale=0.5]{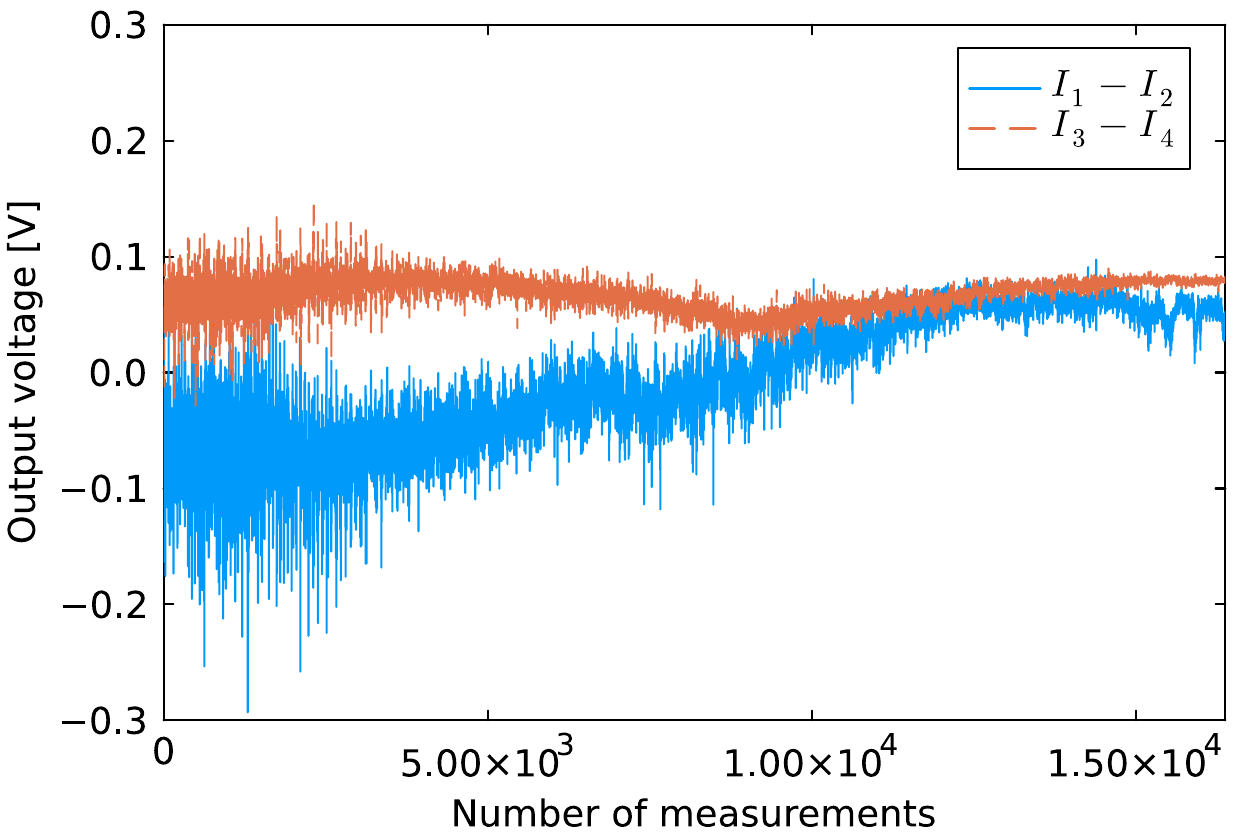}
  \caption{Output voltage of the balanced detectors.}
  \label{fig:intensity}
\end{figure}

Figure \ref{fig:wavefront} shows the wavefront measured without inserting a sample in the optical path of the object light. For these measurements, we used 1,024 orthogonal patterns with a resolution of $32\times32$, enlarged 32 times, generated by the method described in Section 2.1. We also used 4,096 orthogonal patterns with a resolution of $64\times64$, enlarged 16 times. The sampling frequency of the A/D conversion process was \SI{100}{kHz}. The sampled data were averaged every 1,000 results to suppress the influence of interference fluctuations, and the results confirm that the spherical phase derived from the aberration of the optical system can be measured.
The measured wavefront obtained using a microlens array (Thorlabs MLA300-14AR-M) as the sample is shown in Fig. \ref{fig:microlens}. Here, $128\times128$ spatial orthogonal bases, enlarged three times, were used for the measurement. The sampling frequency of the A/D conversion process was \SI{500}{kHz}, and the sampled data were averaged every 10,000 results to obtain a single measurement result. The results confirm that the phase difference originating from the shape of the microlens can be measured. Compared with other methods, e.g., the DH method, which batch measures complex amplitude distributions in two dimensions, the results contains more noise overall.
The outputs of the balanced detectors are shown in Fig. \ref{fig:intensity}. As can be seen, the amplitude is relatively large even in the high spatial frequency region, which is considered to be due to the interference fluctuations caused by the flicker of the SLM and the vibration of the optics. In addition, the baseline of the output voltage is not constant and fluctuates slowly with time. Note that DGH and SPDH both require longer measurement times than the DH method; thus, such fluctuations can be caused by changes in the longitudinal mode of the laser due to changes in temperature and other factors. It is evident that laser stabilization is essential in the DGH and SPDH methods.

\section{Conclusion}
This paper has proposed a method that enables wavefront measurement with the same number of measurements as the conventional GI method by applying DH using parallel phase-shift optics to GI. In the proposed method, four interferograms with a phase shift in steps of $\pi/2$ are measured simultaneously using balanced detectors, and the obtained coefficients are utilized to reconstruct the complex amplitude of the measurement target as a linear combination of the reference light patterns.

The effectiveness of the proposed method was verified experimentally, and the results demonstrated that the wavefront can be reconstructed. In these experiments, we measured the phase difference due to the microlens structure in the demonstration experiment using a microlens array. The proposed method can measure wavefronts using single-pixel detectors; thus, we expect that it can be applied to high-sensitivity measurements and wavelength bands other than visible light. However, noise due to interference fluctuations and the time variation of the laser's longitudinal mode is clearly an issue. In DGH and SPDH, including the proposed method, effective stabilization of both the optics and the laser is an important factor that determines the quality of the measurement.

\section*{CRediT authorship contribution statement}
\textbf{Shuhei Yoshida:} Conceptualization, Data curation, Formal analysis, Investigation, Methodology, Resources, Software, Visualization Writing - original draft.

\section*{Declaration of competing interest}
The author declares that they have no known competing financial interests or personal relationships that could have appeared to influence the work reported in this paper.

\section*{Data availability.}
The Hadamard pattern generation, display program, and wavefront reconstruction program used in this study are available from Ref. \cite{GitHub}.

\bibliographystyle{elsarticle-num}

\end{document}